\documentclass[sigconf,nonacm]{acmart}

\usepackage{xcolor}

\usepackage{amsmath}
\usepackage{relsize}
\usepackage{graphicx}
\usepackage{tikz}
\usepackage{microtype}

\usepackage{listings}
\usepackage[normalem]{ulem} 
\definecolor{lightgray}{rgb}{.95,.95,.95}
\definecolor{darkgray}{rgb}{.4,.4,.4}
\definecolor{purple}{rgb}{0.65, 0.12, 0.82}

\lstdefinelanguage{JavaScript}{
	keywords={typeof, new, true, false, catch, function, return, null, catch, switch, var, if, in, while, do, else, case, break},
	keywordstyle=\color{blue}\bfseries,
	ndkeywords={class, export, boolean, throw, implements, import, this},
	ndkeywordstyle=\color{darkgray}\bfseries,
	identifierstyle=\color{black},
	sensitive=false,
	comment=[l]{//},
	morecomment=[s]{/*}{*/},
	commentstyle=\color{purple}\ttfamily,
	stringstyle=\color{red}\ttfamily,
	morestring=[b]',
	morestring=[b]"
}

\usepackage[capitalize]{cleveref}
\crefname{section}{Sect.}{sections}
\Crefname{section}{Section}{Sections}

\newcommand\parhead[1]{\vspace{1mm}\noindent\textbf{{#1}}}

\newenvironment{definition}
  {\par\medskip\noindent\textbf{Definition.}\itshape}
  {\par\medskip}

\definecolor{rubblue}{RGB}{0,53,96}
\definecolor{rubgray}{RGB}{111,111,111}
\definecolor{rublightgray}{RGB}{245,245,245}
\definecolor{rubgreen}{RGB}{0,120,90}
\definecolor{rubred}{RGB}{200,0,0}
\definecolor{rublightblue}{RGB}{204,229,255}

\newcommand{\summarybox}[3]{\vspace{.5em}\noindent\begin{tikzpicture}
  \node[align=center,draw,thin,minimum width=\columnwidth,inner sep=2.2mm] (titlebox)%
  {\parbox{0.95\columnwidth}{\looseness=-1\noindent\textit{#2\vspace{0.1cm}}}};
  \node[label=left:{\colorbox{white}{\small #1}}] (W) at (titlebox.south east) {};%
  \end{tikzpicture}\vspace{-15pt}}

\begin{document}
	
	
	\title{The Security Feature Location Problem}
	
\author{Kevin Hermann}
\affiliation{%
    \institution{Ruhr University Bochum}
    \city{Bochum}
    \country{Germany}
}

\author{Sven Peldszus}
\affiliation{%
    \institution{Chalmers | University of Gothenburg}
    \city{Gothenburg}
    \country{Sweden}
}

\author{Thorsten Berger}
\affiliation{%
    \institution{Ruhr University Bochum}
    \city{Bochum}
    \country{Germany}
}
\affiliation{%
    \institution{Chalmers | University of Gothenburg}
    \city{Gothenburg}
    \country{Sweden}
}

\author{Adam Shostack}
\affiliation{%
    \institution{University of Washington}
    \city{Seattle}
    \country{USA}
}
\affiliation{%
    \institution{Shostack + Associates}
    \city{Seattle}
    \country{USA}
}

\begin{abstract}
Software security must be realized through security features such as authentication and encryption, but which features does a system implement, and where? We present security feature location: 
the task of relating code locations to security features, enabling developers to understand security implementations and assess whether intended security properties are enforced.
\end{abstract}

\maketitle

\section{Introduction}

\noindent
Security features are concrete code-level mechanisms that serve as essential building blocks of secure software systems\,\cite{Hermann2025Taxonomy,Hermann2025Engineering,ben2014secfeature,mcgraw2004software,Tsipenyuk2005}.
Typical security features, such as authentication, authorization, cryptographic protection, key management, session handling, and incident logging must be correctly designed, implemented, configured, and maintained to enforce security properties, such as confidentiality, integrity, and availability.
Many security vulnerabilities arise when security features are absent, incomplete, or incorrectly implemented.

\begin{figure*}[t]
	\centering
	\includegraphics[width=\textwidth]{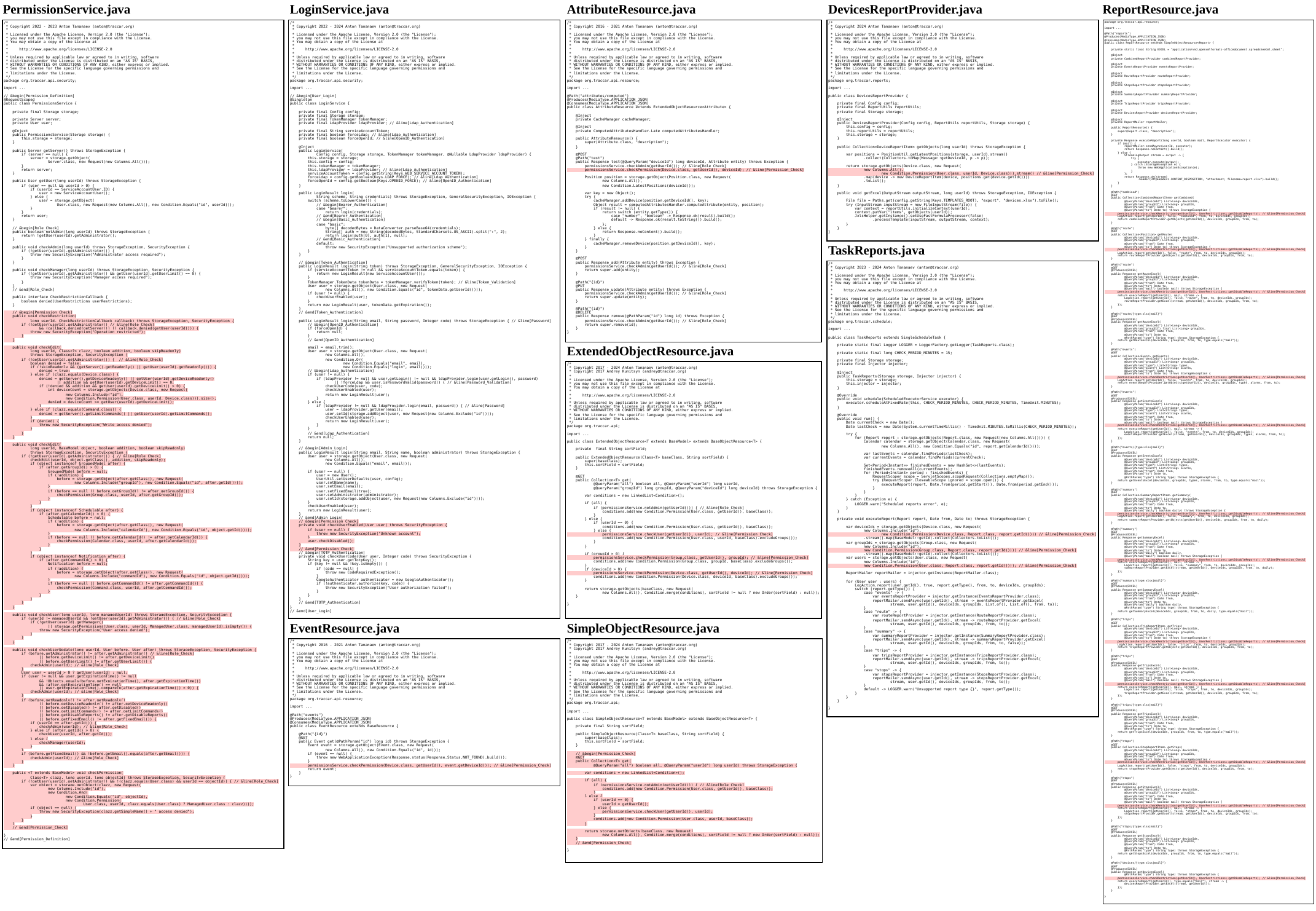}
	\caption{Code of the Traccar GPS tracking software; code responsible for permission checks is highlighted}
	\label{fig:session_expiration}
	\vspace{-.4cm}
\end{figure*}

\looseness=-1
Security features are rarely confined to a single method, class, or library.
\Cref{fig:session_expiration} illustrates their cross-cutting nature by showing how permission checks are distributed in the source code of Traccar, an open-source GPS tracking system.
These checks enforce access control over sensitive operations, including retrieving location data and accessing resources, such as reports.
Permission checks must be performed in all code paths that lead to sensitive operations, which are scattered across the codebase, and often not easy to identify.
A check alone does not provide enough information to determine whether access control is consistently enforced.

\looseness=-1
Even when using third-party libraries to realize security features, the mere presence of a vulnerable library does not imply exploitability. A system is only exploitable if it invokes the affected security feature in a way that triggers the vulnerability.
For example, CVE-2022-23540 in \textit{jsonwebtoken} affects JSON Web Token (JWT) verification: omitting an explicit algorithm definition in a call of \texttt{jwt.verify()} together with passing a false key can lead to bypassing the signature validation. 
Assessing exploitability requires locating not only the vulnerable library, but also the code that invokes it and the code that influences the arguments passed to it, as illustrated below.

\begin{lstlisting}[ language=JavaScript,
	backgroundcolor=\color{lightgray},
	extendedchars=true,
	basicstyle=\scriptsize\ttfamily,
	showstringspaces=false,
	showspaces=false,
	numbers=left,
	numberstyle=\tiny,
	numbersep=3pt,
	tabsize=2,
	breaklines=true,
	showtabs=false,
	captionpos=b,
	label=lst:jsonwebtoken]
	// may or may not return ''undefined''
	const key = getVerificationKey();
	// vulnerable: key may be false + no algorithms
	jwt.verify(token, key);        
	// also problematic: false key + no algorithms
	jwt.verify(token, "", { });          
	// safer: explicit acceptable algorithm
	jwt.verify(token, key, {algorithms: ["RS256"]});
\end{lstlisting}

Both examples raise questions that code does not answer directly: \textit{Is access control enforced on every path? Is the vulnerable library invoked in an exploitable way?}
Such questions are posed over security properties, but must be answered through implementation in code.
However, a single statement is too narrow, since the call to \texttt{jwt.verify()} does not reveal whether the key is valid or invalid.
Likewise, common higher-level abstractions, such as a software bill of materials (SBOM), are too broad, as they only record the presence of \textit{jsonwebtoken}, but not how it is used.
Answering questions on security properties requires an intermediate level---security features---that gathers the code, configuration, and invocation context that together realize one security mechanism.

\looseness=-1
Consider current and upcoming security regulations, which increasingly mandate how software is built and maintained.
The Cyber Resilience Act, for instance, obligates manufacturers to handle vulnerabilities throughout the intended lifespans of their products and to report exploited ones within days.
An SBOM names the affected dependency, but not whether the vulnerable mechanism is invoked, how it is configured, or whether it is reachable, which determines whether the product is affected and what must be fixed.
Consequently, engineers have to locate the security features in their products.

Locating security features, however, is hard.
It requires maintaining an overview over the codebase, which is difficult, especially in large and long-lived software systems.
Over time, functionality is added or modified, the software is reused through cloning, branching, or forking, and development teams change.
When the required knowledge is not recorded, developers need to recover security features and their locations in code, which is laborious and error-prone, especially when developers are not familiar with the codebase\,\cite{Rubin2013FeatureLocation}.
They may miss security-critical code, leaving the system vulnerable to attacks.


We refer to this challenge as the \textbf{security feature location problem:} \textit{given a software system, determine which security features it implements and where the corresponding code is located.}
Solving this problem has direct security impact: it enables developers to assess vulnerability-specific conditions, such as those of CVE-2022-23540, check compliance with security policies and standards, and modify relevant code securely, avoiding unintended changes to security features while supporting deliberate migrations such as from quantum-vulnerable cryptography to post-quantum alternatives.

\looseness=-1
In the age of AI, this challenge is amplified, since software is increasingly written by AI coding agents.
Agents produce and change security-relevant code faster than it can be reviewed, and agents can easily remove parts or whole security features.
The less code developers write, the more they depend on locating security features to assess their correctness and completeness.
Yet, LLMs have a considerable potential to address this challenge.
Agents could record security features as they are written.
LLMs could also recognize security-relevant logic that no API signature reveals.
Explicitly considering and representing security features is an open challenge for researchers and tool builders, especially while agentic practices are still forming and unstructured. 

\looseness=-1
In the remainder, we introduce and discuss the security feature location problem. We describe the notion of security feature, drawing on work in software engineering\,\cite{biggerstaff.ea:1993:featurelocation}, feature-oriented development\,\cite{berger.ea:2015:feature}, and secure software engineering\,\cite{mcgraw2004software,Hermann2025Taxonomy}. We then define the security feature location problem, illustrate its relevance with examples, and outline research challenges towards novel methods and tools to address this problem. 

\section{Software, Features, and Security}
Software is commonly described in terms of the features it provides. When discussing software security, however, systems are often characterized by the properties they must enforce, such as confidentiality, integrity, and availability. These abstract properties must be implemented in code through concrete implementation mechanisms called security features.


\subsection{Software Features}
\noindent
A feature is a label that abstractly represents \mbox{code\,\cite{berger.ea:2015:feature}} and that can be seen as a unit of end-to-end functionality or behavior. Software development processes, such as SCRUM or other agile methods, typically use the notion of feature to plan and manage software systems. Companies strive towards feature teams as opposed to component teams to shorten lead time and increase release frequency to become more agile. As such, features connect requirements, implementation, and maintenance. Developers reason about systems in terms of features when functionality is added, changed, removed, or assessed\,\cite{berger.ea:2015:feature,liu2005modeling}.
To evolve and maintain features, they must understand and locate the implementation that realizes them, known as the feature location\,\cite{Rubin2013FeatureLocation} or concept assignment problem\,\cite{biggerstaff.ea:1993:featurelocation}.

\subsection{Software Security}
Security engineering typically comprises defining the desired security properties of a system, such as confidentiality, integrity, or availability. Such properties need to be refined into project-specific expectations about system design and behavior\,\cite{mcgraw2004software}, often in the form of security requirements.

Methods such as secure design and threat modeling express such expectations in terms of components, trust boundaries, attacker assumptions, and permitted operations\,\cite{Shostack2014ThreatModeling,Basin2006}. Security principles further guide their realization in system design and code. For example, the principle complete mediation calls for permission checks on every relevant access path, whereas the principle least privilege limits the permissions granted to principals, and defense in depth calls for complementary layers of protection\,\cite{Saltzer1975Principles}.

Ultimately, such security expectations are realized through concrete security features in code\,\cite{Hermann2025Taxonomy,Hermann2025Engineering,ben2014secfeature,mcgraw2004software,Tsipenyuk2005}, whose placement, configuration, and interaction across a system affect their intended security effect.


\subsection{Security Features}
Building on the security engineering literature and the general notion of feature in software systems\,\cite{berger.ea:2015:feature}, security features can be characterized as follows.

\begin{definition}
A security feature is a label that represents a concrete security implementation technique in the codebase, spanning artifacts such as source code, configuration files, or deployment descriptors.
\end{definition}

A security feature has a name and is characterized by (i) an intent, such as ``only authenticated users may access resource \emph{X},'' (ii) implementation artifacts, such as code fragments, configuration entries, or infrastructure scripts, and (iii) assumptions about the environment and other features on which it relies.
Examples include access control mechanisms, cryptographic data protection, and key management\,\cite{ben2014secfeature,Tsipenyuk2005}.
Security features may also be composed of sub-features\,\cite{Hermann2025Taxonomy}: for example, access control may combine password-based login and session management.

A security feature represents functionality whose absence would prevent, weaken, or undermine the enforcement of a security property\,\cite{Tsipenyuk2005} by introducing a vulnerability or weakness.
It contributes to enforcing one or more high-level security properties\,\cite{Hermann2025Taxonomy}.
For example, a property such as ``only authorized users can update location data'' may depend on authentication, authorization, identity propagation, and logging.
Such properties emerge from the interaction of multiple features\,\cite{nhlabatsi2008feature}.
For example, authentication establishes an identity, identity propagation carries it through the system, authorization evaluates it against a policy, and logging records security-relevant actions.
Assessing a security property therefore requires locating the relevant security features, their implementation, and their interactions.

A library, algorithm, or platform mechanism is not, by itself, a security feature. It becomes part of one through its specific embedding in code, its configuration, invocation, surrounding logic, dependencies, and assumptions about other mechanisms and the environment\,\cite{ben2014secfeature,Hermann2025Engineering}. As such, security features provide an abstraction necessary to assess the correct implementation of security in software systems.
Security features shift attention from verifying individual components, such as a cryptographic library or access-control routine, in isolation to understanding how security mechanisms are embedded, configured, and combined across a system. Their locations and interactions provide the context and evidence needed to assess whether security properties are enforced.

\section{What is Security Feature Location?}
\looseness=-1
\textit{``What security features are present in my software system? Where are they implemented in code, configuration, and deployment artifacts? How do they interact with other features? Do they collectively enforce the intended security properties?''} These questions illustrate the need to know what security features exist and where they are implemented in a software system, as a prerequisite for assessing, maintaining, or improving its security.
%
%
%
Security feature location is the task of identifying these features in code, specifically:
%
\begin{definition}
Security feature location relates a security feature to the set of locations that implement it. 
A security feature may either be known in advance and need only be located, or first need to be identified within the system’s implementation.
\end{definition}

\looseness=-1
Security feature location inherits challenges from classic feature location\,\cite{biggerstaff.ea:1993:featurelocation,Rubin2013FeatureLocation}, but security features also have distinctive properties. Like classic features, they may be scattered and tangled with each other as well as with functional code. Unlike classic features, however, they are often tied to security-specific concepts such as permissions, identities, secrets, trust boundaries, and attacker assumptions. These concepts are realized through recognizable mechanisms, including security APIs, configuration options, and framework extension points\,\cite{Hermann2025Engineering,Hermann2025Taxonomy}. 
This combination makes the problem challenging, because the security role of a location is not always obvious from code alone, but it also suggests that security-specific cues can support automation more directly than in classic feature location.

Security feature locations may include files, classes, functions, statements, configuration blocks, deployment descriptors, and other implementation artifacts. 
Even libraries might contain security-relevant aspects that could be described in terms of security features.
Security feature location may also describe whether a location is exclusive to one security feature or shared by several security features, the role it plays in enforcing or supporting a security property, the confidence with which it was identified, and its relationships to other feature locations.


Different security tasks may require the locations of a security feature at different levels of granularity.
For example, a developer conducting a security review might need to identify all lines of code where permission checks are implemented.
However, during an incident response, they might only need to identify the files or functions involved in a security breach.
Granularity also applies to the security features themselves. 
A broad assessment may only require locating a coarse-grained feature, such as access control, whereas a focused analysis may require distinguishing sub-features such as authentication and authorization, or even more specific mechanisms such as password-based or multifactor authentication. 
These different views of security features allow developers to reason about security in a modular way, focusing on the relevant security features and their locations for their tasks rather than being overwhelmed by the entire codebase.


Security feature location can be performed using two strategies: proactive and retroactive location.

\textbf{Proactive location} documents security features and their locations as the code is created.
For example, teams may mark the \texttt{jwt.verify()} wrappers and permission checks that belong to a security feature using code annotations, feature flags, or a feature database.
Such recorded locations can be precise when available, but they are often incomplete or absent in practice, decay as the system evolves, and require ongoing maintenance.

\textbf{Retroactive location} recovers security features and their locations from the implementation itself, either manually or with tool support, such as information retrieval and static or dynamic analysis. 
Unlike proactive location, this strategy does not require previously recorded location information, but it can be laborious and may produce imprecise or incomplete results.

\begin{figure}
	\centering
	\resizebox{\linewidth}{!}{\input{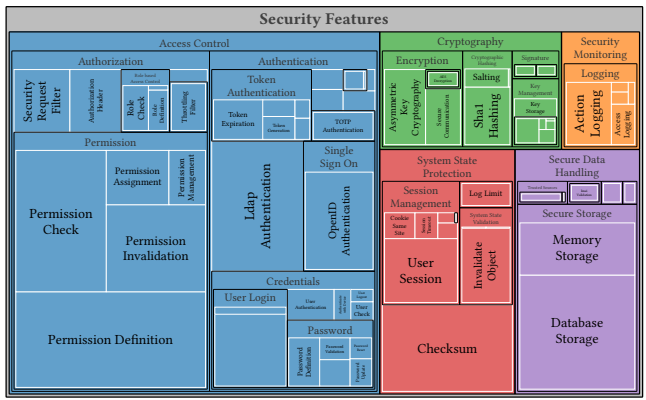}}
	\caption{Relative size of security features in the Traccar GPS system.}
	\label{fig:tree_map}
\end{figure}

\begin{figure}
	\centering
	\resizebox{1\linewidth}{!}{\input{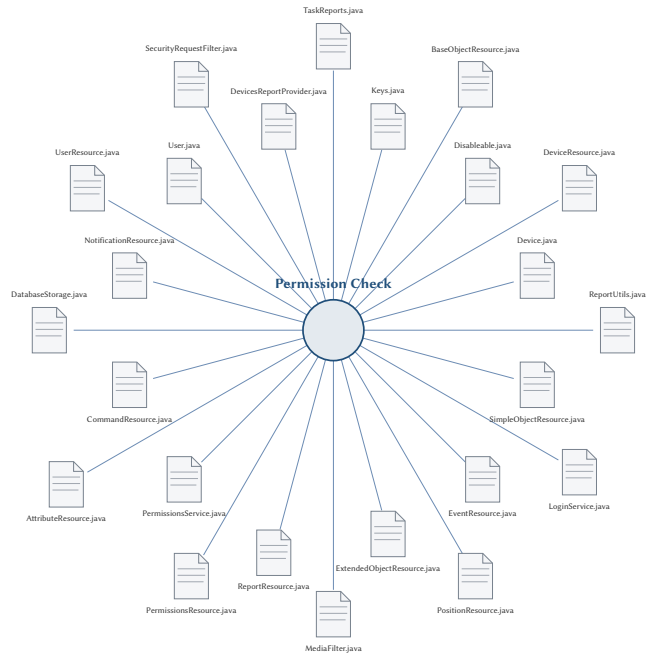}}
	\caption{Scattering of permission checks in the Traccar GPS system.}
	\label{fig:scattering}
\end{figure}

\begin{figure*}[t]
	\centering
	\resizebox{\linewidth}{!}{\input{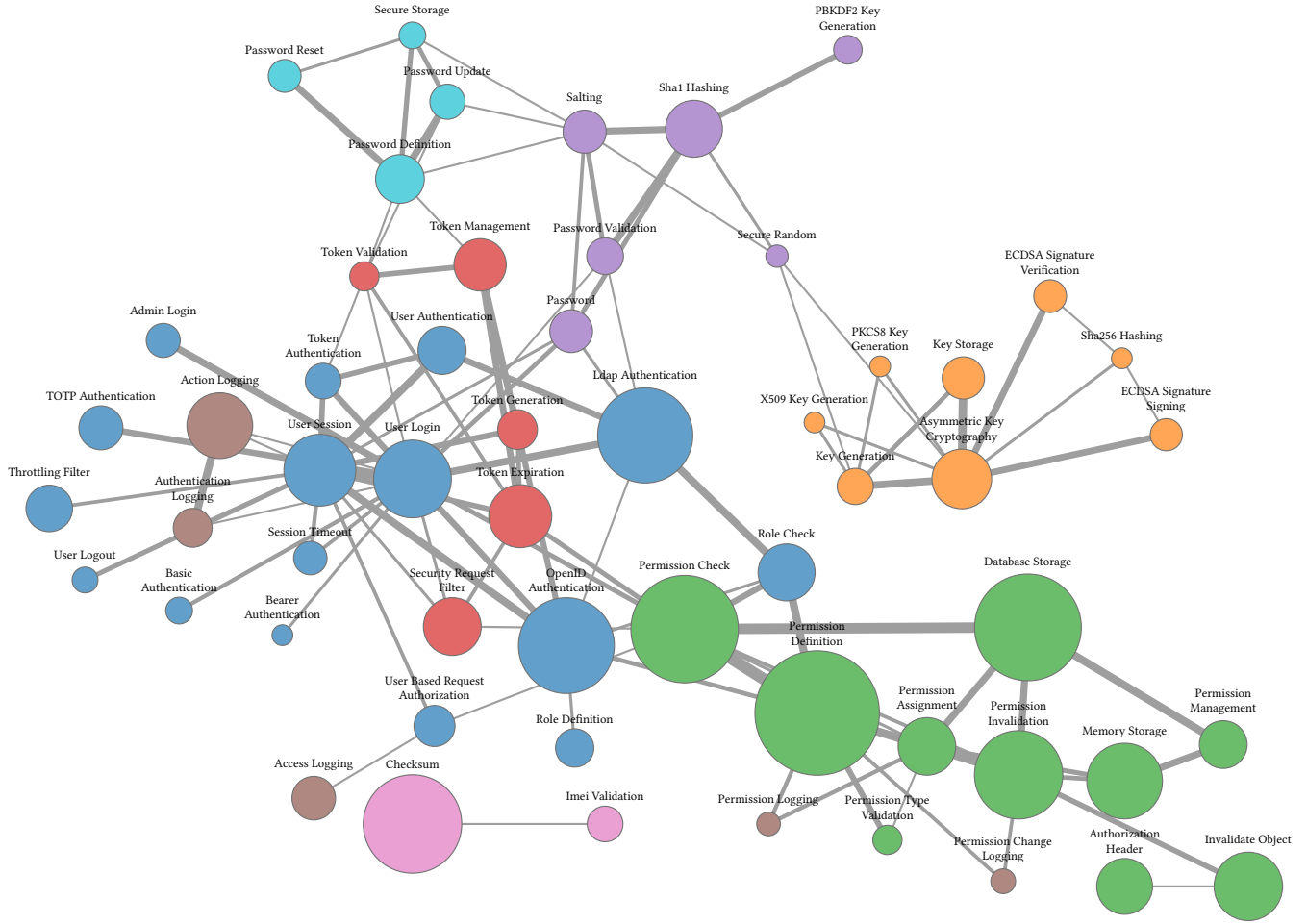}}
	\caption{Tangling and interaction between security features in the Traccar GPS system. Each circle is a security feature, its area is proportional to the feature's size (annotated lines of code), and the thickness of a connecting line grows with the number of code lines in which the two features are tangled. Circles with the same color belong to the same security feature category.}
	\label{fig:tangling_cluster}
\end{figure*}

\section{What Can Security Feature Location Enable?}
\noindent
Security feature location helps to relate design-level concepts, such as security properties, threat assumptions, and security principles, to concrete code-level artifacts, such as API calls, configuration entries, and policy rules.
This allows developers to reason about security in a modularized way and obtain code-level evidence for security claims, focusing on relevant features and their locations without being overwhelmed by the entire codebase.
As such, security feature location is a prerequisite for many security-relevant tasks, including security reviews, compliance assessment, incident response, and secure software evolution.
Beyond these tasks, security feature locations can also provide security-specific context to improve automated security analyses.
%
%



\parhead{Security Reviews and Compliance Assessment.}
\looseness=-1
Security reviews require confidence that all relevant security mechanisms have been identified and examined.
Developers may know that a security property should hold, but lack the code-level evidence needed to assess whether it is actually enforced.
Security feature location provides them with this code-level evidence to perform security reviews and assess compliance with security standards.
In particular, it helps developers determine whether authorization is consistently enforced, whether sensitive data is protected along relevant execution paths, whether a disclosed vulnerability is reachable, and which locations must be modified when a security mechanism changes.

Missing locations of security features, such as permission checks, which are often subtle and scattered, may cause reviewers to overlook critical vulnerabilities.
Security feature location provides reviewers with an overview of security-relevant code, helping them focus their effort and increasing the coverage of manual review.
Automated support for locating security features, potentially combined with control-flow, policy, or coverage analysis, can further highlight vulnerabilities or inconsistencies in the implementation, such as permission checks that are applied in some but not all code paths leading to a sensitive operation.

\parhead{Incident Response and Analysis.}
Responding to security incidents and newly discovered threats is often time-critical.
Developers must rapidly identify the relevant security features and their implementation locations to assess the situation and determine whether remediation is necessary.
When a security-relevant issue is observed, such as unexpected or unauthorized access to a resource, developers may need to trace the relevant permission checks, authentication mechanisms, and access-control data to understand its cause, assess whether it has been exploited, and determine what must be fixed.
In other situations, new external security information may trigger the analysis. 
For example, when a vulnerability in a library, such as the one in \textit{jsonwebtoken}, is disclosed, developers must first determine whether their system uses the affected security feature in a vulnerable way.
If so, they must then identify the code that needs to be changed. 
Both kinds of analyses require knowledge of the relationships among security features and their implementation locations. 
Making this information explicitly available can accelerate the analysis, help developers understand how a security issue relates to existing security mechanisms, and identify the necessary changes to prevent exploitation or recurrence.

\parhead{Secure Software Evolution.}
Software systems continuously evolve.
Changes to functionality, dependencies, or configuration may invalidate security assumptions or cause noncompliance with security policies or standards.
During evolution, a system may remain functionally correct while becoming vulnerable because changes invalidate security assumptions or properties, or introduce new attack vectors\,\cite{Hermann2025Engineering}.
Making security features and their locations explicit helps developers prevent accidental changes to security-critical code when evolving a system.

Security feature location also supports impact analysis and regression checks by identifying the code and configuration artifacts relevant to a change. 
For example, when migrating from an elliptic-curve-based cryptographic mechanism to a post-quantum alternative, developers need to identify all locations in which the affected mechanism is used.
This may require considering parameters and configuration in addition to API calls, since the relevant cryptographic operation or concrete algorithm may not be evident from the call itself.
The same Bouncy Castle \texttt{init()} call, for instance, can realize encryption or decryption depending on a mode parameter, while concrete algorithms may be specified separately in configuration files\,\cite{Hermann2025Taxonomy}.
Similarly, when tightening access-control policies, developers can use the locations of permission checks to identify those that depend on changed roles or attributes.

Security feature location further supports migration to new standards or platforms.
Porting a system to a new identity provider, encryption standard, or logging infrastructure requires identifying all code and configuration that currently implement the relevant security features.
Without support, such migrations are error-prone and may leave vulnerable legacy mechanisms behind.




\parhead{Enhancing Static Analysis.}
Beyond directly supporting developers in security tasks, security feature location can improve code-level security analysis by connecting security expectations to implementation locations. 
Prior work shows that security analysis can be optimized when information about security features is available, for example, 
to check whether planned mechanisms are present in code, or whether confidential data is actually encrypted as specified in a threat model\,\cite{Tuma2023Compliance}.
It can also help triage static-analysis findings by identifying which files, methods, or configuration fragments implement security-critical functionality\,\cite{Peldszus2026}.
More broadly, security feature location can provide entry points or scopes for analyses and tests that otherwise struggle with scale and precision, including information-flow analysis, access-control checks, and fuzzing of paths that interact with security features.

\section{Why is Security Feature Location a Problem?}

We illustrate the security feature location problem using the open-source GPS tracking system Traccar.
\Cref{fig:tree_map} shows its security features, their relative sizes, and their division into five categories: access control (blue), cryptography (green), secure data handling (purple), system state protection (red), and security monitoring (orange).
Overall, Traccar's codebase comprises 77,051 lines of Java code, of which 4,702 (6.1\%) contribute to security features.

Recall the feature permission check from \Cref{fig:session_expiration}.
The feature is scattered across 66 distinct locations and 24 files, illustrated in \Cref{fig:scattering}.
For example, when a user attempts to access a resource, such as a GPS device or a location history, the code must check whether the user is authenticated and holds the required permissions.
Although the checks themselves are relatively small, they are widely scattered and easy to miss, and being able to locate them is important for developers to know whether access control policies are correctly enforced, or to change them when needed.

Recovering these permission checks could be done by developers as follows.
First, they need to identify an entry point, such as the usage of a security library or a configuration parameter.
Such an entry point may be found through documentation, knowledge of the codebase, or by using information retrieval techniques to search for keywords in code, configuration, and deployment artifacts.
From there, they have to trace the control flow to identify surrounding code, functions, classes, or other artifacts that contribute to the security feature.
They then need to identify where these artifacts are used in the system and repeat the process until they have recovered the feature's implementation to the extent the security-related task requires.

This process, however, is laborious and error-prone.
Documentation of security features is often incomplete or outdated, development teams change often, and information retrieval techniques return too many false positives or miss important locations to be useful in practice.
The lack of effective techniques puts developers, who are rarely security experts\,\cite{Hermann2025Engineering}, in a difficult position: \textit{they lack a clear understanding of how security features are realized in the system}, which makes it hard for them to assess whether security properties are consistently enforced, or where vulnerabilities may arise.

Security features are not only scattered across many locations, but they are also tangled with one another, as illustrated in \Cref{fig:tangling_cluster}.
Although they primarily interact with security features within their own category, such as cryptography, authentication, or authorization, they frequently also interact with security features in other categories. For example, cryptographic operations are used to hash user passwords, permission checks depend on the authentication state, and security logging is used across all features.
This forces developers in yet another difficult position: \textit{they must understand how security features interact with one another}, as making changes to one feature may affect the security properties of another\,\cite{nhlabatsi2008feature}. 
For example, changing the password hashing algorithm to comply with new security standards may affect the security of authentication in many ways: it requires updates to the login workflow to use the new algorithm, password reset functionality to re-hash existing passwords, and the authentication state to be invalidated for all users until the migration is complete.

Scattering and interaction also mean that there is no single view of a security feature that is appropriate for every security task. As discussed above, a security review may require a broad view of permission checks to identify missing or inconsistent checks. This view includes their implementation locations, the operations they protect, and the policies they enforce. By contrast, incident response may require a narrower view, focusing on the permission checks and interacting security features involved in a particular unauthorized access. Supporting such task-specific views is challenging because security feature location must determine not only the relevant locations, but also the appropriate level of granularity for the task, which may only become clear as the analysis proceeds.

\summarybox{Why Security Feature Location is a Problem}{Security features are often realized at \textbf{multiple levels of granularity}, from low-level permission checks and cryptographic operations to high-level access control policies.
Implementations are both \textbf{scattered} (spread across many files and components) and \textbf{tangled} (interleaved with one another).
As software evolves over time, development teams change, and knowledge about security features fades, security feature location becomes increasingly difficult.}{1cm}



\section{Research Challenges}
%
So far, we presented the security feature location problem conceptually and illustrated its relevance and opportunities with concrete examples.
But how can we actually solve it?
Despite its importance and opportunities for automation, security feature location has received little attention as a problem in its own right.
The following research challenges outline what is needed to turn the security feature location problem into practical methods and tools for developers and security engineers and to make systems more secure.

\subsection{Nature of Security Features}
Before we can locate security features, we need to improve our empirical understanding of them in real software systems. That is, we need to determine syntactic and semantic characteristics, as well as their relation to other common security abstractions, such as security properties.

\parhead{C1.1: What constitutes a security feature?}
Security features can be based on a wide range of security mechanisms, including authentication, authorization, session handling, incident logging, and data protection.
However, they are typically used in combination with non-security features.
For example, while session timeouts may be used to prevent session hijacking, they also serve to free up resources in a web application.
But where does functionality, which provides value to stakeholders in software systems, end and where does security begin?

\looseness=-1
Studies should investigate the nature of security features, how they realize security properties, how security features can be distinguished from non-security functionality, and how their boundaries can be characterized.

\parhead{C1.2: What are the characteristics of security features?}
Security features may be implemented through code, configuration, or deployment artifacts.
They are often cross-cutting, scattered across multiple locations, and may be realized through different implementation techniques using different frameworks or libraries.
To understand how security features enforce security properties and how they can be located, we need to characterize how they manifest in real systems. Which implementation characteristics indicate the presence of a security feature, and how do these characteristics relate to the security properties it helps enforce?

Studies should identify recurring implementation patterns, such as security APIs, configuration entries, data flows, or deployment descriptors, characterize them across systems, and investigate how they contribute to enforcing security properties. 
Such knowledge can reveal which cues are useful for locating security features and understanding their role in enforcing security properties.

\parhead{C1.3: How do security features interact?}
To enforce security properties, security features often interact with each other\,\cite{nhlabatsi2008feature}.
For example, a permission check depends on authentication to establish the identity on which an authorization decision is based. 
Such interactions can determine whether a security property is actually enforced. 
Incorrect, incomplete, or unexpected interactions between security features may weaken security properties or introduce vulnerabilities even when the individual features are implemented correctly. 
Security features may also interact with functional features, for example when permission checks guard security-sensitive operations. 
But how do security features interact, how do these interactions affect security properties, and how can they be captured?

Studies should characterize recurring interaction patterns between security features across systems and investigate how they contribute to, weaken, or violate security properties. 
Understanding these patterns can support locating related security features and identifying interactions that may introduce vulnerabilities.

\subsection{Methods and Tools}
With a better empirical understanding of security features, effective methods and tools can be built.

\parhead{C2.1: How can we represent security features?}
Security features provide an abstraction between security properties, threat models, and security principles on the one hand and concrete code-level implementations on the other.
The former are typically expressed at design time, whereas the latter evolve through code, configuration, and dependencies during implementation and evolution.
Security properties rarely map one-to-one to security features, but emerge from the composition of multiple security features.
But how do we connect these design-level security properties to the concrete artifacts that implement them?

A research challenge is to design a representation of security features as a means to connect security requirements, assumptions, and guarantees to their concrete implementations within a system.
Such a representation should also characterize the security role of a location, such as checking permissions, validating input, propagating identity, encrypting data, or logging a security event.

\parhead{C2.2: How can we proactively record and maintain security feature locations?}
Security-relevant information is often available when a security feature is introduced or modified. 
During development, engineers may know the security purpose of a change, the security feature it affects, and the implementation artifacts used to realize it. 
Instead of losing this information and recovering it later, techniques could proactively record and continuously maintain information about security features and their implementation locations. 
Recorded security feature locations can in turn serve as input to subsequent development activities, informing developers and tools about which artifacts are security-relevant and what features they contribute to.

Agentic development may provide an additional opportunity for proactive recording. 
Security-related prompts already express the intent of a change, while coding agents have access to the artifacts they modify to realize it. 
This information could be used to record and update security feature locations during development, while existing feature locations could in turn be provided as context for subsequent changes, helping agents account for and preserve existing security mechanisms as the system evolves.

A central challenge is deciding at what granularity security feature locations should be recorded. This concerns both the granularity of the security features themselves and the granularity of their implementation locations. 
Recording only coarse-grained features or file-level locations may be inexpensive but insufficient for later development or security tasks, while continuously recording fine-grained sub-features or individual statements, configuration values, and interactions may impose unnecessary overhead and be difficult to maintain.
But how can security features be recorded with little additional developer effort, and at what levels of granularity in the security feature hierarchy (e.g., cryptography, encryption, AES) and in the code structure (e.g., statements, lines, functions, classes, namespaces)?

Studies should investigate techniques that derive, use, and update security feature locations throughout development and security engineering. 
This includes determining how information about security features can be captured with little additional effort during conventional development, how coding agents can use existing security feature information as context and update it automatically, how appropriate feature and location granularity can be determined, and how recorded locations can be kept consistent with the evolving implementation.

\parhead{C2.3: How can we retroactively obtain the security feature locations needed for security tasks?}
Proactively recorded security feature locations may provide a useful baseline, but they may not contain all information required for a particular security task. 
Incident response may require tracing the locations and interactions surrounding a particular permission check, while assessing a newly disclosed vulnerability may require identifying uses of specific APIs, parameters, or configuration values associated with a specific security feature or sub-feature.
Such information may be too detailed or task-specific to maintain continuously, especially when future security tasks cannot be anticipated. Retroactive recovery is also necessary for existing systems for which no security feature information has been recorded.

Security features are often implemented using API calls and configuration parameters of security frameworks and libraries\,\cite{Hermann2025Taxonomy,Hermann2025Engineering}. 
Such cues can provide entry points from which relevant code locations can be traced through the system.
However, security features are rarely confined to recognizable entry points and may be scattered across code and configuration or realized partly or entirely through custom application code. 
The challenge is to determine which artifacts jointly constitute a security feature, how far its implementation extends beyond identifiable entry points, and at what granularity it needs to be recovered for a particular security task. 
Moreover, when recovering feature locations from scratch, it may be difficult to determine whether all relevant locations have been found.
But how can we determine how far a security feature's implementation extends, and when all of its locations have been found?

Existing feature-location research remains general-purpose and does not address security-specific concerns\,\cite{biggerstaff.ea:1993:featurelocation,Rubin2013FeatureLocation}.
Studies should investigate techniques such as static and dynamic analysis, information retrieval, and LLM-based approaches for recovering security feature locations. 
In particular, studies should investigate to what extent LLMs can identify security-relevant code from its semantics and surrounding context, including custom implementations for which no known security API provides an entry point.
LLM-based techniques could also be combined with security-specific cues and program analyses to identify and propagate feature locations.
Existing proactively recorded locations can provide additional input, allowing such techniques to refine the existing mapping or recover finer-grained and task-specific information when needed. 
Studies should investigate how these techniques can be combined and evaluate their precision, recall, scalability, and ability to recover sufficiently complete feature locations for concrete security tasks.



\parhead{C2.4: How can we contextualize security feature locations for security tasks?}
Identifying security feature locations alone does not reveal how these locations work together or how interactions between security features affect the enforcement of security properties. 
As illustrated in \Cref{fig:tree_map,fig:scattering,fig:tangling_cluster}, existing feature-location and visualization techniques provide a step in this direction by exposing properties such as scattering, tangling, and relationships between feature locations. 
However, such representations are not necessarily tailored to security engineering. 
For security tasks, developers may need additional context about the security role of particular locations, the security properties they contribute to, the assumptions they rely on, and the relationships among interacting features. 
But which aspects of this security-specific context are relevant to a particular task, and how should it be made accessible to developers?

Future work should investigate how representations and visualizations of feature locations can be contextualized for security tasks. 
This includes determining which information about security properties, assumptions, and interactions should be exposed for a particular task and how developers can navigate between this context and the corresponding implementation locations.
Such techniques should enable developers to interpret the security relevance of scattered and tangled implementations without requiring them to manually reconstruct this context.

%

\subsection{Overhead and Benefits} 
Once we have developed methods and tools for security feature location, we need to investigate to what extent they increase the security of a system and whether the benefits outweigh the overhead they impose.

\parhead{C3.1: What are the benefits of security feature location?}
The value of security feature location ultimately depends on whether its results improve concrete security tasks.
We need evaluation methods that measure its impact on tasks such as security review, incident response, and compliance assessment.
But what are appropriate evaluation methods and metrics for such tasks, and how can we assess whether improvements in task performance translate into better security outcomes?

\looseness=-1
Security feature location research needs to develop task-specific evaluation methods and metrics, as well as ground-truth datasets of security feature locations in real systems.
Evaluations should establish whether and under which conditions security feature location leads to better security-relevant decisions and outcomes.

\parhead{C3.2: What is the overhead of locating security features?}
Although locating security features can help developers understand and maintain a system's security, it also incurs overhead.
Developers may need to learn new tools, adapt their workflows, maintain proactively recorded feature locations, or inspect results produced by retroactive techniques.
We need to understand the overhead of security feature location in terms of time, effort, and cognitive load.
But how can we measure the overhead of security feature location?

Studies should therefore evaluate not only the accuracy and effectiveness of security feature location techniques, but also the costs they impose on developers and security engineers.
Understanding these costs is essential for assessing their practical usefulness and likelihood of adoption.

\parhead{C3.3: What level of granularity provides the best trade-off for different security tasks?}
Different security tasks may require reasoning about security features at different levels of granularity.
A security review may require a fine-grained understanding of how security features enforce security properties, whereas some secure software evolution tasks may require only a coarse-grained understanding of which artifacts realize a security feature.
Moreover, the same task may be performed at different levels of granularity, creating a trade-off between effort and effectiveness.
A finer-grained representation may provide more precise guidance but require more effort to recover, maintain, and understand, while a coarser-grained representation may be cheaper to obtain but insufficient for some security decisions. 
But what level of granularity provides the best balance for a given security task, and how does this affect the overhead and benefits of security feature location?

Studies with practitioners should compare different granularities for the same security tasks and measure their effects on task performance and security-relevant outcomes. 
They should also investigate when maintaining fine-grained security feature location information continuously provides sufficient benefit to justify its overhead and when additional detail can instead be recovered on demand.

\section{Conclusion}
\noindent
Effective security review, maintenance, and evolution require knowing which security features a system implements and where they are located.
We advocate making the notion of security features more explicit and argue that security features are useful abstractions for security engineering.
We defined security feature location as the task of identifying the security features a system implements and where they reside in its implementation. We discussed why it is a problem in practice. Specifically, security features realize security properties in software, but the knowledge on their implementation is quickly lost in a large and evolving codebase.

\looseness=-1
Security feature location connects design-level security expectations to the concrete artifacts that realize them. This connection helps developers to assess whether security properties are enforced and to securely change the relevant code without accidentally weakening it.
Establishing this connection requires that practitioners explicitly consider, represent, and maintain security features and that researchers develop effective methods and tools for security feature location.



\section{Acknowledgements}
The authors used GPT (OpenAI) and Claude Opus/Sonnet (Anthropic) models for language-related support, including proofreading and clarifying arguments. The models were also used to assist in the visualization of data for Figures 2--4. All content was reviewed by the authors, who take full responsibility for the manuscript.





\bibliographystyle{ACM-Reference-Format}
\bibliography{literature}
\end{document}